# Cutting Mechanics of Soft Compressible Solids –
# Force-radius scaling versus bulk modulus

Bharath Antarvedi Goda[a], Mattia Bacca[a*]

[a]*Mechanical Engineering Department, Institute of Applied Mathematics, School of Biomedical Engineering, University of British Columbia, Vancouver BC V6T1Z4, Canada*

[*]Corresponding author. *E-mail address:* mbacca@mech.ubc.ca

## Abstract

Cutting mechanics in soft solids present a complex mechanical challenge due to the intricate behavior of soft ductile materials as they undergo crack nucleation and propagation. Recent research has explored the relationship between the cutting force needed to continuously cut a soft material and the radius of the wire (blade). A typical simplifying assumption is that of material incompressibility, albeit no material in nature is really incompressible. In this study, we relax this assumption and examine how material (in)compressibility influences the correlation between cutting forces and material properties like toughness and modulus. The ratio $\mu/\kappa$, where $\mu$ and $\kappa$ are the shear and bulk moduli, indicates the material's degree of compressibility, where incompressible materials have $\mu/\kappa = 0$, and larger $\mu/\kappa$ provide higher volumetric compressibility. Following previous observations, we obtain two cutting regimes: (i) high toughness or small wire radius, and (ii) low toughness or large wire radius. Regime (i) is dominated by frictional dissipation, while regime (ii) is dominated by adhesive debonding and/or the wear resistance of the material. These regimes are controlled by the ratio between the wire radius and the elasto-cohesive length of the material: the critical crack opening displacement at fast fracture under uniaxial tension. In the large radius, regime (ii), our theoretical findings reveal that incompressible materials require larger forces. Notably, however, the elasto-cohesive length of the material, defining the transition wire radius between regimes (i) and (ii), is larger for compressible materials, which are therefore more likely to be cut in regime (i), and thus requiring larger cutting forces.

## 1. Introduction

Cutting mechanics in soft solids has gained significant attention owing to its extensive range of engineering and biomedical applications, from food processing (Xu2022) and manufacturing (Williams2010) to robotic surgery (Brodie2018) just to name a few. As a consequence of entropy-driven elasticity, soft elastomers undergo large deformations prior to failure (Lake1967). Thus, fracture mechanics in these materials must be analyzed with the use of nonlinear elasticity (Long2021). Over the past few decades, significant attention has been paid to understanding soft-material failure under remote loading conditions such as in the case of pure shear and uniaxial tension (Rivlin1953, Thomas1955, Long2021). Cutting mechanics, however, represents a more complex physical scenario due to the presence of friction and a more complex stress distribution in the specimen. Moreover, while traditional fracture only concerned with the geometry of the sample and that of the crack, in cutting mechanics the geometry of the blade must be accounted for too. Hence, the importance of defining 'sharpness'. Sharpness not only influences the stress concentrations at the nucleated crack but also the extent of the development of friction, which appears to control the magnitude of cutting forces. The first cutting tests were performed by (Lake1978) using a Y-shaped specimen pulled at both ends by a force, while the tip of the blade was in contact with the center of the 'y'. This complex apparatus was developed with the aim of decoupling stress concentration from friction. The same apparatus was adopted by (Zhang2019, Zhang2021) to further investigate the role of friction as the ratio between the tip radius of the blade and two characteristic length scales of the material, namely, the size of the fracture process zone (FPZ) and the size of the nonlinear elastic zone (Long2021). However, while friction was purposefully minimized in this apparatus, its role in defining the magnitude of cutting forces should be thoroughly quantified. To fully investigate cutting, including friction, one should minimize the geometric complexity of the blade. Hence, a good candidate is wire cutting, where the blade is composed of a rigid cylinder (wire) of radius $R_w$, which constitutes the only geometric variable controlling sharpness. (Kamyab1998) investigated wire cutting mechanics in cheese, and obtained the cutting force as linearly proportional to $R_w$, where the force at $R_w = 0$ is only given by the toughness of the sample, while the force-radius slope is given by the product of the friction coefficient between the wire and the sample and the yield strength of the sample. More recently, such a force-radius linear correlation we also observed by (Zhang2021, Püffel2023, Goda2024) both theoretically and experimentally. Particularly, (Goda2024) quantified the distinct contribution of friction and the wear resistance of the sample, with the latter being attributed to damage evolution in the elastomer (unlike yielding in Kamyab1998). The same study also highlighted that the cutting force at $R_w = 0$ corresponds to cleavage toughness, *i.e.*, the sole dissipative contribution of the energy required to break the links crossing the crack plane. Cleavage toughness provides the minimum fracture energy at the net of the dissipative phenomena generated in the FPZ. At $R_w > 0$, the linear force-radius proportionality is given by friction and wear resistance, with the latter providing the FPZ contribution to toughness.

A common assumption in all the above models is that of incompressibility of the cut sample. In this study we relax this assumption and explore the effect of bulk modulus on cutting forces. Following the approach proposed by (Goda2024), we quantify the contributions to cutting force given by toughness, friction and wear, as a function of wire radius and the elastic response of the material. Our study confirms previous findings on incompressible materials (Goda2024), showing a small-radius or high-toughness regime (i), where cutting forces are given by Coulomb friction, and a larger-radius or low-toughness regime (ii), where cutting forces are given by the wear strength of the specimen. We also find that bulk modulus affects the transition radius between regimes (i) and (ii), where incompressible materials are more likely to fall in (ii) while highly compressible ones will fall in (i). In regime (ii) we observe that incompressible materials require larger cutting forces, while the opposite occurs in regime (i). The latter is also in line with previous investigations on puncture mechanics (Fregonese2023).

## 2. Model

The model system is reported in Figure 1. Here, a specimen of soft compressible hyperelastic material of length $L$, height $H$, and thickness $T$, has a pre-existing cut of depth $c$ from the free surface, along the direction $X_1$. The inserted rigid cylindrical wire has cross-sectional radius $R_w$. Based on (Goda2024), the work of cutting per unit depth $D$ in direction $X_1$, at steady-state, is

$$F_{ss}dD = \Gamma_o T dD + dQ_d \tag{1}$$

where $\Gamma_o$ is the cleavage toughness and provides the minimum energy required to advance the crack surface by the unit $TdD$, and $Q_d$ is the energy dissipated by additional dissipative mechanisms other than fracture. These are friction, adhesive debonding between specimen and wire, and surface damage in the specimen (wear) (Goda2024). Eq. (1) can also be rewritten as

$$F_{ss} = \Gamma_o T + F_d \qquad (2)$$

where $F_d = dQ_d/dD$ is the force contribution due to the abovementioned dissipative mechanisms.

From Eq. (1), in the absence of dissipative mechanisms other than fracture, $dQ_d = 0$, and thus $F_d = 0$. From Eq. (2) we then have the minimum steady-state cutting force as $F_{ss} = \Gamma_o T$. This minimum force is independent of the wire radius and the elastic response of the material. As shown by (Kamyab1998), (Zhang2019), (Zhang2021), (Püffel2023), and (Goda2024), this minimum force corresponds to a wire of zero radius, $R_w = 0$. As shown by (Goda2024), in this ideal case the mechanical work of the penetrating zero-radius wire corresponds to the sole energy required to break the bonds in the crack (cut) plane, hence $\Gamma_o$ is cleavage toughness, and the crack does not develop a process zone. A wire having a radius that is larger than zero will unavoidably trigger additional dissipative mechanisms, which we attribute to friction, adhesive debonding and surface damage (wear), generating $F_d > 0$. As in (Goda2024), we quantify $F_d$ via integration of the interfacial shear strength $\tau_c$ (resisting specimen-wire sliding) across the contact surface and projected along the wire direction $X_1$ (Figure 2a), giving

$$F_d = 2R_w T \int_{\theta_1}^{\theta_2} \tau_c \sin\theta \, d\theta \qquad (3)$$

Here, $\theta_1$ and $\theta_2$ are the contact angles shown in Figure 2a, and

$$\tau_c = \xi P + \tau_w \qquad (4)$$

where $\xi$ is the friction coefficient, $P$ is the contact pressure at the wire-specimen interface (so $\xi P$ is Coulomb friction), and $\tau_w$ is the pressure-independent friction, which we attribute to adhesive debonding or surface damage (wear) (Fregonese2022, Goda2024). As can be deduced from Eq. (3) and (4), $F_d$ depends on the interfacial properties $\xi$ and $\tau_w$, as well as the configurational variables $\theta_1$, $\theta_2$, and $P$, which depend on the competition between toughness and the elastic deformation of the specimen, as well as the radius of the wire $R_w$. The elastic deformation is described as follows.

We assume the problem is plane strain, and analyze half model, thanks to the symmetry of the system with respect to the $X_1$ axis. Via finite element analysis (FEA), we evaluate the $J$ integral around the crack tip. FEA is performed using ABAQUS, where the wire is modeled using R2D2 elements while the specimen is modeled using plane-strain CPE4H elements.

The specimen's hyperelastic response is described via Ogden's (Ogden1972) strain energy density (SED)

$$W = W_d + W_v \qquad (5a)$$

additively composed by the deviatoric strain energy density

$$W_d = \frac{2\mu}{\alpha^2}(\bar{I}_\alpha - 3) \qquad (5b)$$

with

$$\bar{I}_\alpha = \bar{\lambda}_1^\alpha + \bar{\lambda}_2^\alpha + \bar{\lambda}_3^\alpha \tag{5c}$$

and the volumetric strain energy density

$$W_v = \frac{K}{2}(\Delta - 1)^2 \tag{5c}$$

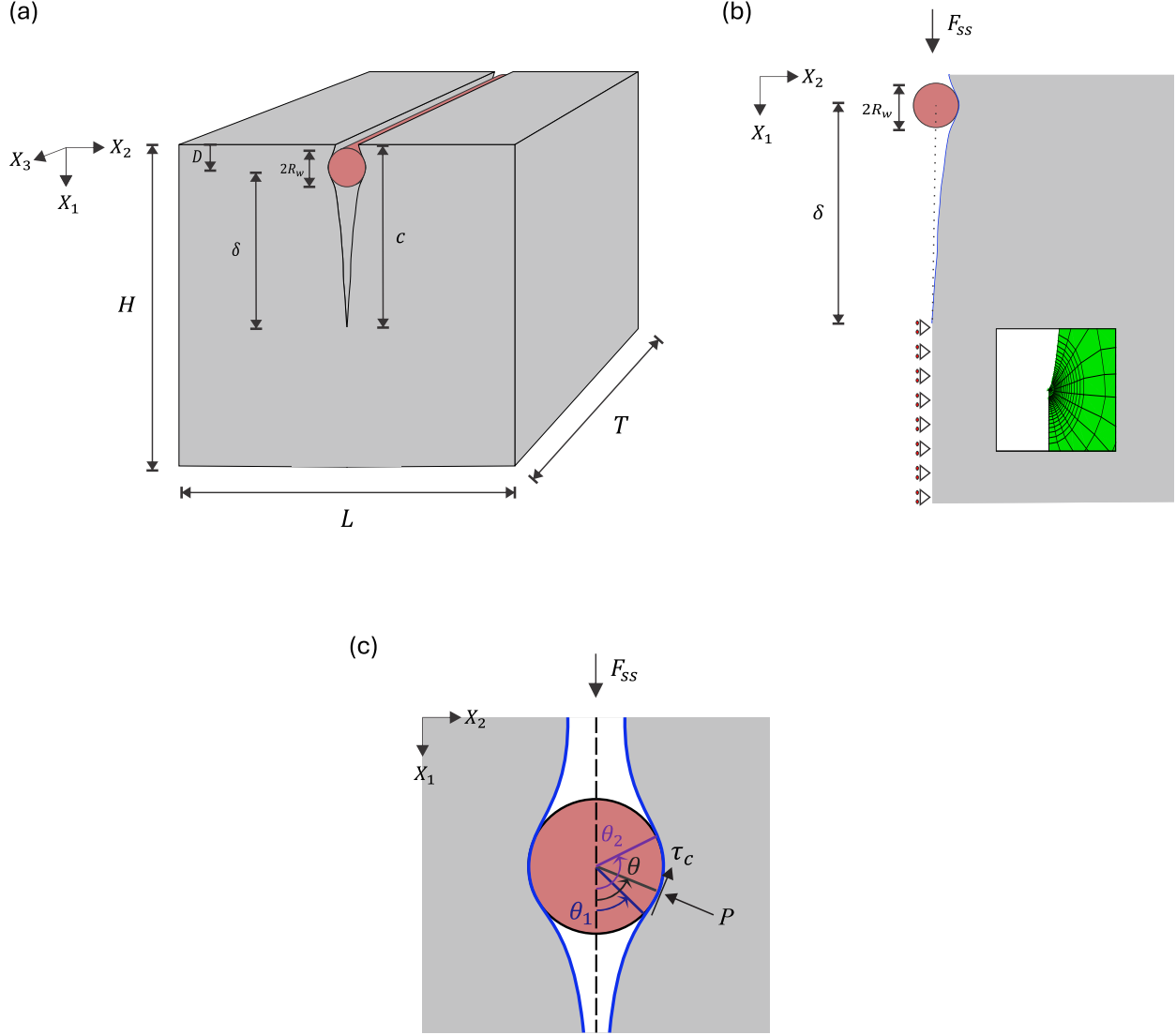

**Figure 1**: (a) Schematics of wire cutting of a soft compressible hyperelastic specimen (grey) with a pre-existing crack. The wire is described as a rigid cylinder (brown). (b) 2D Plane-Strain half model (snippet showing mesh around the tip), thanks to symmetry about the $X_1$-axis. The material is described via an Ogden hyperelastic, and we used finite element analysis (FEA) to evaluate the energy release rate, $J$, as a function of the distance $\delta$ between the wire and the crack tip, assuming the wire is inserted into the cut at a sufficient distance from the free surface ($D \gg R_w$). (c) Schematic of the interfacial stress distribution, for the force balance in Eq. (3).

Here, $\mu$ is the shear modulus, $K$ is the bulk modulus, $\bar{\lambda}_i = \lambda_i J^{-1/3}$ is the $i$-th deviatoric principal stretch, with $\lambda_i$ the corresponding total principal stretch and $\Delta = dv/dv_0$ the volumetric swelling ratio, where $dv$ and $dv_0$ are the unit volumes in current and reference configuration, respectively. Now, the principal Cauchy stress in direction 1, is given by

$$\sigma_1 = \frac{\lambda_1}{\Delta} \frac{\partial W}{\partial \lambda_1} \tag{6a}$$

which, from Eq. (5) gives

$$\sigma_1 = \sigma_{d_1} - p \tag{6b}$$

where

$$\sigma_{d_1} = \frac{2\mu}{\alpha \Delta}\left(\bar{\lambda}_1^\alpha - \frac{\bar{I}_\alpha}{3}\right) \tag{6c}$$

is the deviatoric component of $\sigma_1$, and

$$p = -K(\Delta - 1) \tag{6d}$$

is the hydrostatic pressure $p = -(\sigma_1 + \sigma_2 + \sigma_3)/3$. We can now write both SED and stress in dimensionless form as

$$\frac{W}{\mu} = \frac{2}{\alpha^2}(\bar{I}_\alpha - 3) + \frac{1}{2}\frac{K}{\mu}(\Delta - 1)^2 \tag{7}$$

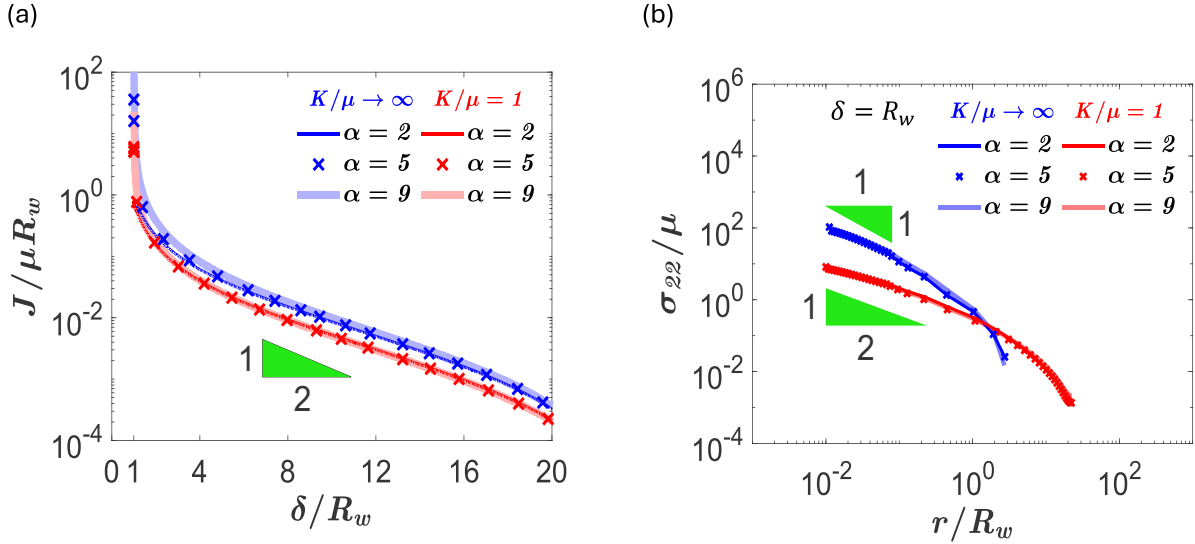

**Figure 2**: (a) Semi-log plot of the dimensionless energy release rate $J/\mu R_w$ versus dimensionless distance $\delta/R_w$ between wire and crack tip. (b) Log-log plot of the dimensionless Cauchy stress $\sigma_{22}/\mu$ versus the dimensionless distance from the crack tip $r/R_w$ (with $r = X_1 - c$ in Figure 1) at $\delta = R_w$. We assume incompressibility ratios of $K/\mu \to \infty$ (blue) and $K/\mu = 1$ (red) with strain stiffening parameters $\alpha = 2$, $\alpha = 5$, and $\alpha = 9$.

and

$$\frac{\sigma_1}{\mu} = \frac{2}{\alpha \Delta}\left(\bar{\lambda}_1^\alpha - \frac{\bar{I}_\alpha}{3}\right) + \frac{K}{\mu}(\Delta - 1) \tag{8}$$

from Eq. (5) and (6), respectively. Here, both dimensionless quantities only depend on two dimensionless elastic parameters, namely, the Ogden strain-stiffening parameter $\alpha$ and the incompressibility ratio $K/\mu$.

## 3. Results and Discussion

In Figure 2a we report the dimensionless energy release rate $J/\mu R_w$ versus dimensionless distance $\delta/R_w$ between the wire and the crack tip calculated from FEA for an incompressible material ($K/\mu \to \infty$) (blue) (Goda2024) and a highly compressible material ($K/\mu = 1$) (red). For large tip-wire distances, $\delta \gg R_w$, we observe the simple relation

$$\frac{J}{\mu R_w} \approx \beta \left(\frac{R_w}{\delta}\right)^2 \tag{9}$$

In this regime we also observe $J \ll \mu R_w$. Eq. (9) applies for any strain stiffening $\alpha$ and incompressibility ratio $K/\mu$, where the fitting coefficient $\beta$ generally depends on $\alpha$ and $K/\mu$. This regime is also called 'far field' cutting due to the long wire-tip distance and is opposed to that of 'indentation cutting' (Lake1978, Zhang2019, Goda2024), occurring for $\delta \approx R_w$ and giving $J \gg \mu R_w$. Notably, the dimensionless constant $\beta$ in Eq. (9) is independent of $\alpha$, thus, only depending on $K/\mu$. Also, $\beta$, and thus $J$ (for given $\mu$, $\delta$, and $R_w$), is proportional to $K/\mu$, with an almost 2-fold increase from $K/\mu = 1$ to $K/\mu \to \infty$. In Table 1 we report the values of $\beta$ at various values of $K/\mu$, averaged over the explored values of $\alpha$ (2,5,9). Given the proportionality between $\beta$ and $K/\mu$ we can deduce that incompressibility increases the energy release rate, if all other quantities are fixed; or, most importantly, for a given toughness $\Gamma_0$ (assuming $J = \Gamma_0$) we have that a compressible material will admit shorter wire-tip distance $\delta$. This is simply due to the volumetric absorption of elastic deformation allowed in the material, while an incompressible material will have overall more rigidity and move the crack tip to a further distance from the wire.

| $\kappa/\mu$ | $\infty$ | 30 | 5 | 2 | 1 |
|---|---|---|---|---|---|
| $\beta$ | 2.02 | 1.62 | 1.33 | 1.21 | 1 |

**Table 1**: Calculated values of the fitting coefficient $\beta$, from Eq. (9), as a function of $K/\mu$. $\beta$ does not significantly change with respect to $\alpha$ and these values are averaged within the explored $\alpha = 2,5,9$.

The concentration of dimensionless tensile stress $\sigma_{22}/\mu$, normal to the crack (cut) plane, in the proximity of the crack tip is shown in Fig 2b, as a function of the dimensionless distance $r/R_w$ from the tip ($r = X_1 - c$). In this figure we describe the indentation-cutting regime, adopting $\delta = R_w$. The near-tip stress concentration can be generally described as

$$\frac{\sigma_{22}}{\mu} \approx \left(\frac{nJ}{\mu \pi R_w}\right)^{1/n} \left(\frac{R_w}{r}\right)^{1/n} \tag{10}$$

where $n = 2$ recovers *linear elastic fracture mechanics* (LEFM) solution (Tada1973), while $n = 1$ recovers *nonlinear elastic fracture mechanics* (NEFM) solution calculated via neo-Hookean elasticity (with $\alpha = 2$) (Long2021). As we can observe from the log-log plot in Figure 2b, in comparison with Eq. (10), a hyperelastic incompressible material ($K/\mu \to \infty$, blue) exhibits a stress concentration that is more akin to that of NEFM, while a highly compressible one ($K/\mu = 1$, red) exhibits a concentration that resembles that obtained from LEFM. Notably, these stress concentrations appear to be nearly independent of $\alpha$. As discussed by (Long2021), LEFM gives stress triaxiality with a concentration of hydrostatic stress, for which $\sigma_{11} = \sigma_{33} = \sigma_{22}$, while NEFM provides stress uniaxiality, so that $\sigma_{11} \approx \sigma_{33} \approx 0$. Thus, we can define the general relation $\sigma_{11} \approx \sigma_{33} \approx \sigma_{22}(n-1)$.

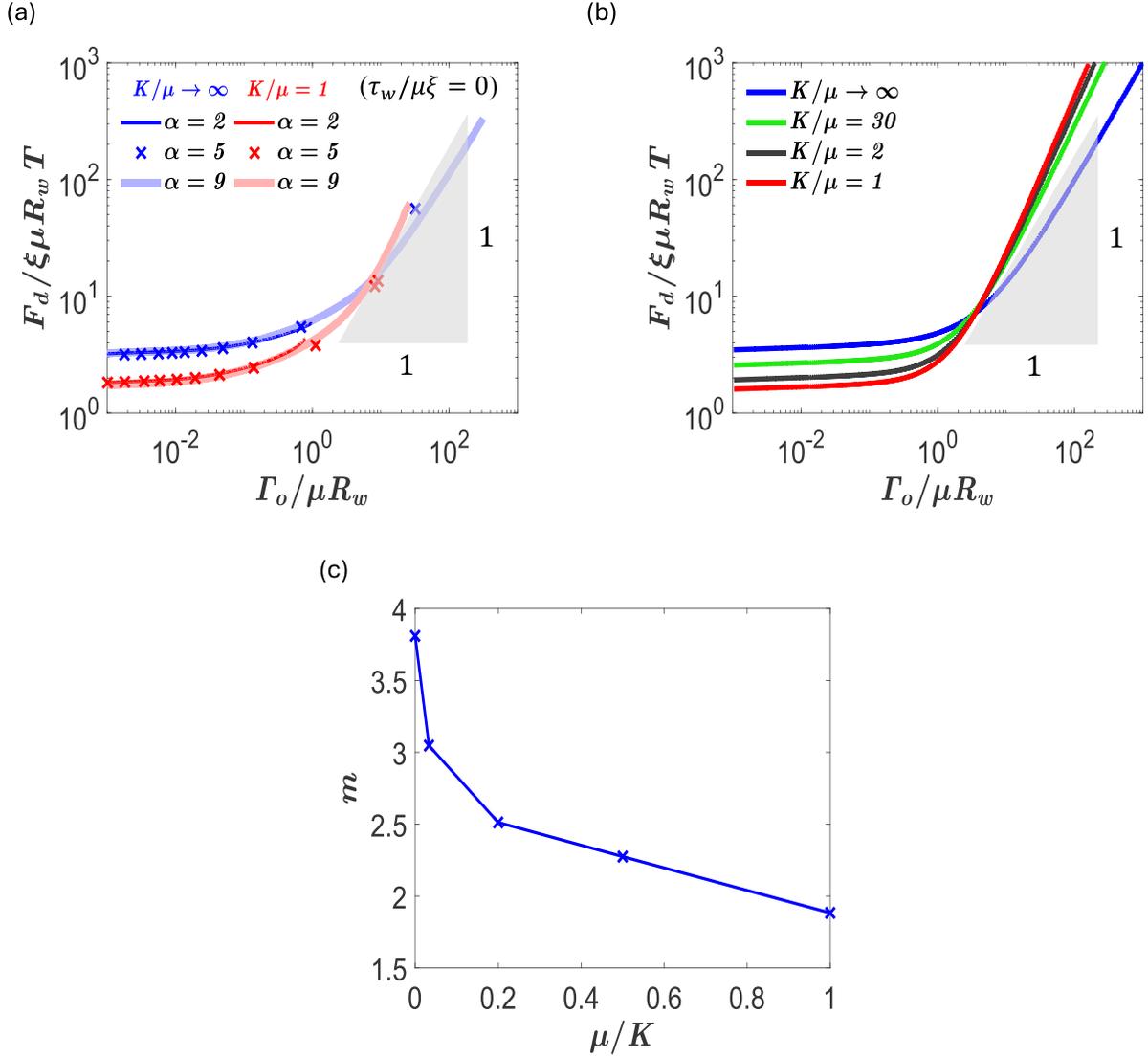

**Figure 3**: (a) Dissipative force contribution $F_d$, from Eq. (3) and in dimensionless form, for strain stiffening coefficients $\alpha = 2, 5, 9$, and compressibility ratios $K/\mu = 1$ (highly compressible) and $\infty$ (incompressible). (b) Same results as in the previous figure, but $\alpha = 9$, and $K/\mu = 1, 2, 30, \infty$. (c) For the low toughness regime, describing the plateau in the previous two figures (plot domain on the left), we report the $m$ coefficient used in Eq. (12) and (13) as a function of $\mu/K$, where $\mu/K = 0$ for incompressible materials.

Consider now the calculated energy release rate $J$ corresponds to the cleavage toughness $\Gamma_0$ of the material ($J = \Gamma_0$), so that the corresponding dimensionless wire-tip distance $\delta/R_w$ provides the critical configuration between wire and specimen. Under this hypothesis we have that $\delta/R_w$ is uniquely defined by the dimensionless parameter $\Gamma_0/\mu R_w$ (Goda2024), which controls whether a given material and wire radius exhibit far-filed cutting ($\delta \gg R_w$) or indentation cutting ($\delta \sim R_w$) conditions. The pressure distribution $P$ in Eq. (4) and the angles of wire-specimen contact $\theta_1$ and $\theta_2$ in Eq. (3) depend on $\delta/R_w$, and thus on $\Gamma_0/\mu R_w$. Thus, we can now integrate all the contact shear stresses from Eq. (4) substituted into (3), and obtain the dimensionless dissipative force $F_d/\mu T R_w$, as reported in Figure 3b-c. In these figures, the dimensionless force is a function of the dimensionless toughness $\Gamma_0/\mu R_w$, the Ogden strain stiffening coefficient $\alpha$, the incompressibility ratio $K/\mu$, the dimensionless wear strength $\tau_w/\mu$, and the friction coefficient $\zeta$. We report $F_d/\mu T R_w \zeta$ in the y-axis of Figure 3b-c for simplicity of exposition of our results as in (Goda2024).

As also observed by (Goda2024), in Figure 3b-c we see two regimes, namely (i) high toughness-to-modulus, or small radius, regime, where $F_d$ is mainly due to friction and corresponds to indentation cutting; (ii) low toughness-to-modulus, or large radius regime, where $F_d$ is mainly due to adhesive debonding or wear, and corresponds to far-field cutting. These two regimes of friction vs adhesive debonding domination were also observed by (Fregonese2022) in puncture mechanics. In both puncture and cutting, friction domination is due to larger contact pressure, while wear or adhesive debonding, described as a pressure-independent frictional resistance (Fregonese2022, Goda2024), dominates at low contact pressure.

In case (i) we have $\Gamma_o \geq \mu R_w$ or $R_w \leq \Gamma_o/\mu$, and the results in Figure 3c can be fitted to (Goda2024)

$$\frac{F_d}{\mu R_w T} \approx \xi \left(\frac{\Gamma_o}{\mu R_w}\right)^p \qquad (11)$$

where the power-law fitting coefficient $p$ is close to unity.

In case (ii), we have $\Gamma_o \leq \mu R_w$ or $R_w \geq \Gamma_o/\mu$, and the results in Figure 3c can be fitted to

$$\frac{F_d}{\mu R_w T} \approx m \left(\xi + \frac{\tau_w}{\mu}\right) \qquad (12)$$

where the fitting coefficient $m$ is $m \approx 4$ for the incompressible case, as obtained by (Goda2024).

Eq. (11) and (12) results from curve fitting our dimensionless results from FEA, with an $R^2 \geq 0.97$.

By merging the above equations and substituting the result into Eq. (2), we have the final dimensionless cutting force

$$\frac{F_{ss}}{\Gamma_o T} \approx 1 + \xi \left(\frac{\mu R_w}{\Gamma_o}\right)^{1-p} + m \left(\xi + \frac{\tau_w}{\mu}\right)\frac{\mu R_w}{\Gamma_o} \qquad (13)$$

where $\Gamma_o T$ is the zero-radius cutting force, requiring only the energy cost of rupture of the bonds located in the crack (cut) plane. In Eq. (13), the wire radius $R_w$ is rescaled by the material length $l_e = \Gamma_o/\mu$. This property, called elasto-cohesive length (Long2021) is proportional to the size of the nonlinear zone around the crack tip, where nonlinear elastic deformations are concentrated, and is also proportional to the critical crack opening displacement in uniaxial tension. Notably, we can also observe that $l_e = \Gamma_o/\mu$, in Eq. (13), provides the transition radius between the horizontal line (far-field cutting regime (ii)) and the inclined line (indentation cutting regime i) in Figure 3b-

c. This transition, however, is notably impacted by material compressibility, where a larger $K/\mu$ yields a larger transition radius, as can be observed in Figure 3c. This suggests that incompressible materials may have a larger nonlinear elastic zone around the crack tip (Long2021), compared to highly compressible materials. This observation might seem in contradiction with Figure 2b, where incompressible materials exhibit NLEFM stress concentration, while compressible ones show LEFM-like behavior; suggesting that the nonlinear zone in compressible materials is smaller than that for incompressible ones. However, we must consider that the size of the nonlinear zone in Figure 2b scales as $J/\mu$, not $\Gamma_o/\mu$, and, as shown in Figure 2a, incompressible materials exhibit larger $J$. To the authors' knowledge, there is currently no study that investigates the relation between bulk modulus and elasto-cohesive length of a material, and our observations might inspire further investigation in this aspect.

Figure 3a-b shows that $p < 1$ for incompressible materials ($K/\mu \to \infty$) and $\alpha = 9$, where the force-radius proportionality in Eq. (13) is ensured also from the second term of the right-hand side. Notably, the same figure also shows that compressible materials, for which $K/\mu \leq 30$, with $\alpha = 9$, exhibit $p > 1$. In this case the second term on the right-hand side of Eq. (13) provides an inverse force-radius proportionality, which might be observed for relatively small radii, i.e., $R_w \ll l_e$. A similar inverse proportionality due to $p > 1$ was also observed in incompressible materials for $\alpha = 2$ and 5 in (Goda2024). The above observations provide an interesting relation between cutting force scaling and the compressibility of a material, but further experimental validation is needed to confirm all these theoretical findings. From this theoretical observation, however, we predict that highly compressible materials exhibit higher cutting resistance at small radii, or large toughness (in regime (i)). The same deduction was also proposed by (Fregonese2023) in puncture mechanics.

At large $R_w/l_e$ (in regime (ii)), assuming negligible friction $\xi$, we have that the second term on the right-hand side of Eq. (13) can be neglected. In this case, Eq. (13) gives a linear proportionality between cutting force and wire radius (Kamyab1998, Zhang2019, Zhang2021, Püffel2023, Goda2024). In the case in which $\tau_w \gg \xi\mu$, i.e., with low friction $\xi$ or high $\tau_w/\mu$ (high adhesion or wear resistance), Eq. (13) becomes

$$\frac{F_{ss}}{\Gamma_o T} \approx 1 + m\frac{\tau_w R_w}{\Gamma_o} \tag{14}$$

The force-radius linear scaling in Eq. (14) shows that now the wire radius $R_w$ is rescaled by another material length, $l_w = \Gamma_o/\tau_w$. (Goda2024) proposed the correlation $\tau_w \sim W_f$, with $W_f$ the work of fracture, i.e., the average critical strain energy density in a pristine specimen at crack nucleation under uniaxial tension. For the studied specimen of polyacrylamide gel, (Goda2024) observed a close correlation so that $\tau_w \approx W_f$. Under this assumption, $l_w \approx l_f$, with $l_f = \Gamma_o/W_f$ the fracto-cohesive length (Long2021), i.e., a material length that is proportional to the size of the fracture process zone (FPZ). The latter hosts most of the dissipative mechanisms that generate toughness in ductile materials in uniaxial tension. Following Eq. (14), the force-radius relation is controlled by the FPZ size in uniaxial tension. Notably, some uniaxial tension experiments evidenced an increase of toughness when the crack was created with an incision of radius $R_w$ (Thomas1955). As discussed in (Goda2024), large radii, for which $R_w > l_f$ are likely to *irreversibly* blunt the crack tip by effectually enlarging the FPZ beyond the maximum theoretical size predicted by $l_f$ via uniaxial tension or pure-shear tests. This can be seen if we take $F_{ss} \approx \Gamma T$, where, from Eq. (14), $\Gamma \approx \Gamma_o + m\tau_w R_w$ or $\Gamma \approx \Gamma_o(1 + mR_w/l_w)$. Here, $mR_w/l_w$ -or $mR_w/l_f$- provides an amplification factor to the total toughness (cleavage plus FPZ contribution) and suggests that the FPZ size in cutting is controlled by the wire radius, hence by sharpness. We should, however, consider that the correlation $l_w \approx l_f$ observed in (Goda2024) might be challenged in highly compressible materials.

From Figure 3a-c we can generally observe that both fitting coefficients $p$ and $m$ in Eq. (11) and (12) depend on the Ogden strain stiffening coefficient $\alpha$ and the incompressibility ratio $K/\mu$. For $\alpha = 2$, recovering neo-Hookean elasticity, element distortion prevented us from developing all the results until the high-toughness regime in Figure 3b. However, we can deduce that both $p$ (the slope of the tendency line on the right-hand side of the domain) and $m$ (the offset of the horizontal line on the left-hand side of the domain) are mildly affected by the strain stiffening coefficient $\alpha$, while significantly affected by the incompressibility ratio $K/\mu$.

In Figure 3c we show the relation between the fitting coefficient $m$, used in Eq. (12), (13), and (14), and the ratio $\mu/K$. Incompressible materials, where $\mu/K = 0$ ($K/\mu \to \infty$) exhibit $m \approx 4$, confirming previous observations (Goda2024). Compressible materials having thus larger $\mu/K$ (smaller $K/\mu$) exhibit a progressive reduction of $m$ with the minimum observed $m \approx 1.9$ at $\mu = K$ (highly compressible material).

From all the above observations we can conclude that incompressible materials, compared to highly compressible ones, are more likely to be cut in a far-field cutting regime where dissipative mechanisms resisting cutting are mainly given by pressure-independent sliding resistance such as surface damage (wear). In the far-field cutting regime, incompressible materials also require larger cutting forces due to the higher value of $m$ reported in Figure 3c. Highly compressible materials therefore require smaller cutting forces in the far-field cutting regime. However, as observed in Figure 3a-b, highly compressible materials are more likely to be cut in the indentation cutting regime (regime (i)), i.e. in the small radius or high toughness regime in which they might exhibit inverse proportionality between cutting force and wire radius. In this case, compressible materials (with low bulk modulus) will exhibit higher cutting resistance compared to their incompressible counterparts, as also theoretically observed in puncture resistance (Fregonese2023).

Our study describes volumetric compressibility with the linear elastic volumetric strain energy described in Eq. (5c). As materials might undergo significant volumetric compression during cutting, the required volumetric strain energy is likely to increase more than what is predicted in this equation. Our study should therefore be taken as the first step.

Finally, our study only focuses on the steady-state cutting force, after the cut has already been initiated. Much like in puncture mechanics (Fregonese2021), the insertion of the blade involves a snap-through instability which is difficult to describe and model mathematically. We expect a significant influence of the bulk modulus on the critical force for blade insertion and leave blade insertion as a future focus.

## 4. Conclusion

Our theoretical study sheds light on the role of volumetric compressibility in soft solid cutting, suggesting that cutting forces are affected by the bulk modulus of the material. The correlation between bulk modulus and cutting forces depends on the other material properties involved, such as toughness, shear modulus, and strain stiffening coefficient, as well as the radius of the wire defining blade sharpness. We observe two cutting regimes, confirming previous findings. These are (i) indentation cutting, where the tip of the blade touches the sample at the crack tip, and (ii) far-field cutting, where the tip of the blade does not touch the crack tip. In regime (i), we find that the cutting forces are higher in highly compressible materials than their incompressible counterparts, suggesting an additional cutting strength emerging from volumetric compressibility. This validates previous findings on puncture mechanics. Conversely, in regime (ii), the cutting forces exerted by highly compressible materials are smaller than their incompressible counterpart. The transition wire radius between regimes (i) and (ii) is given by the ratio of toughness over shear modulus, is called elasto-cohesive length of the material, and corresponds to the critical crack opening displacement at fast fracture in notched samples subjected to uniaxial loading. Notably, we find that this transition length is larger in highly compressible materials. Thus, at a given wire radius, highly compressible materials are more likely to fall in the cutting regime (i), while incompressible materials are more likely to fall in regime (ii).

Our findings highlight the important role of the bulk modulus and the volumetric compressible behavior of the material in defining its cutting resistance in wire-cutting experiments. However, our findings are purely theoretical, and experimental validation is needed. For the latter, researchers will be challenged in their ability to develop materials of a desired bulk modulus without affecting all other properties. Such an endeavor will bring more insight into our study.

**Acknowledgments**

The work was supported by the Human Frontiers Science Program (RGY0073/2020), the Department of National Defense (DND) of Canada (CFPMN1–026), and the Natural Sciences and Engineering Research Council of Canada (NSERC) (RGPIN-2017–04464).

**References**

Brodie, A. and Vasdev, N., 2018. The future of robotic surgery. *The Annals of The Royal College of Surgeons of England*, *100*(Supplement 7), pp.4-13.

Fregonese, S. and Bacca, M., 2021. Piercing soft solids: A mechanical theory for needle insertion. *Journal of the Mechanics and Physics of Solids*, *154*, p.104497.

Fregonese, S. and Bacca, M., 2022. How friction and adhesion affect the mechanics of deep penetration in soft solids. *Soft Matter*, *18*(36), pp.6882-6887.

Fregonese, S., Tong, Z., Wang, S. and Bacca, M., 2023. Theoretical puncture mechanics of soft compressible solids. *Journal of Applied Mechanics*, *90*(11), p.111003.

Goda, B.A., Ma, Z., Fregonese, S. and Bacca, M., 2024. Cutting soft matter: scaling relations controlled by toughness, friction, and wear. *Soft Matter*, *20*(30), pp.6016-6022.

Kamyab, I., Chakrabarti, S. and Williams, J.G., 1998. Cutting cheese with wire. *Journal of Materials Science*, *33*, pp.2763-2770.

Xu, W., Wang, J., Deng, Y., Li, J., Yan, T., Zhao, S., Yang, X., Xu, E., Wang, W. and Liu, D., 2022. Advanced cutting techniques for solid food: Mechanisms, applications, modeling approaches, and future perspectives. *Comprehensive Reviews in Food Science and Food Safety*, *21*(2), pp.1568-1597.

Lake, G.J. and Yeoh, O.H., 1978. Measurement of rubber cutting resistance in the absence of friction. *International Journal of Fracture*, *14*, pp.509-526.

Lake, G.J. and Thomas, A.G., 1967. The strength of highly elastic materials. *Proceedings of the Royal Society of London. Series A. Mathematical and Physical Sciences*, *300*(1460), pp.108-119.

Long, R., Hui, C.Y., Gong, J.P. and Bouchbinder, E., 2021. The fracture of highly deformable soft materials: A tale of two length scales. *Annual Review of Condensed Matter Physics*, *12*(1), pp.71-94.

Ogden, R.W., 1972. Large deformation isotropic elasticity–on the correlation of theory and experiment for incompressible rubberlike solids. *Proceedings of the Royal Society of London. A. Mathematical and Physical Sciences*, *326*(1567), pp.565-584.

Püffel, F., Walthaus, O.K., Kang, V. and Labonte, D., 2023. Biomechanics of cutting: sharpness, wear sensitivity and the scaling of cutting forces in leaf-cutter ant mandibles. *Philosophical Transactions of the Royal Society B*, *378*(1891), p.20220547.

Rivlin, R.S. and Thomas, A.G., 1953. Rupture of rubber. I. Characteristic energy for tearing. *Journal of polymer science*, *10*(3), pp.291-318.

Thomas, A.G., 1955. Rupture of rubber. II. The strain concentration at an incision. *Journal of Polymer science*, *18*(88), pp.177-188.

Williams, J.G., Patel, Y. and Blackman, B.R.K., 2010. A fracture mechanics analysis of cutting and machining. *Engineering Fracture Mechanics*, *77*(2), pp.293-308.

Zhang, B., Shiang, C.S., Yang, S.J. and Hutchens, S.B., 2019. Y-shaped cutting for the systematic characterization of cutting and tearing. *Experimental Mechanics*, *59*, pp.517-529.

Zhang, B. and Hutchens, S.B., 2021. On the relationship between cutting and tearing in soft elastic solids. *Soft Matter*, *17*(28), pp.6728-6741.